\documentstyle[preprint,aps,amssymb,amsfonts]{revtex}
\tightenlines
\def\e{\varepsilon}
\def\ee{\end{equation}}

\def\be{\begin{equation}}
\def\ee{\end{equation}}
\def\Ored{\tilde{\Omega} (E)}
\def\O{\Omega (E,M)}
\def\H3{H_w(E,M;\{S\}_{\cal N})}

\def\D#1#2{\delta_{#1,#2}}
\def\bhme#1{\tilde{N}_{\cal A}^{#1}}
\def\bhm#1#2{N_{\cal A}^{#1,#2}}
\def\average#1{\left\langle #1 \right\rangle}
\def\averagebf#1{{\left\langle #1 \right\rangle}}

\def\micavbf#1#2#3{{\left\langle #1 \right\rangle} (#2,#3)}
\def\konfig{\Gamma_{L^d}}

\begin{document}
\draft
\title{Speeding Up Computer Simulations:\\
The Transition Observable Method
}
\author{M.~Kastner and M.~Promberger}
\address{Institut f\"ur Theoretische Physik, Universit\"at 
Erlangen-N\"urnberg, Staudtstrasse 7, D--91058 Erlangen, Germany
}
\author{J.D.Mu\~noz}
\address{
ICA1 Universit\"at Stuttgart, Pfaffenwaldring 27, D-70569 Stuttgart, Germany\\
Permanent address: Departamento de F\'{\i}sica, Universidad Nacional de Colombia, Bogot\'a D.C., Colombia
}
\date{\today}
\maketitle
\begin{abstract}
A method is presented which allows for a tremendous speed-up of computer simulations of statistical systems by orders of magnitude. This speed-up is achieved by means of a new observable, while the algorithm of the simulation remains unchanged.
\end{abstract}

\narrowtext

\section{Introduction}

In performing a Monte Carlo simulation, one or several
observables are chosen, for which a simulation average is recorded. A common choice for
such an observable is related to a histogram as introduced by Salsburg \cite{Salsburg} and popularized by Swendsen 
and Ferrenberg \cite{SwendFer}, \cite{Swendsen}. The histogram allows for the
estimation of the density of states, 
from which a variety of physically
interesting system properties can be computed. Recently, Oliveira et
al. \cite{Oliveira} suggested a new method, called transition observable method
throughout this paper, which likewise enables an estimation
of the density of states but leads to considerable reduction of the
computing time. In other words, the simulation may be dramatically accelerated
without modifying the algorithm but by changing the observable recorded during
the simulation.%
\footnote{%
Since this new observable represents properties of the system
under consideration, it is of course completely independent of the particular
simulation setup.}

If the standard histogram method is applied to magnetic systems, the number of microstates with
energy $E$ and magnetization $M$ is counted during the course of simulation,
i.e., every microstate yields {\em one} entry in an energy-magnetization
histogram. From this histogram the density of states can be computed (cf.~Sec.~\ref{histotech}).\\
In the transition observable method, each microstate of the Monte Carlo sample
is exploited in a much more sophisticated way. For the particular realization
of the method introduced in Sec.~\ref{comparison}, this means: In an extended
histogram, the number of possible transitions is recorded from particular
microstates (in the Monte Carlo sample) with
energy $E$ and magnetization $M$ to "neighbouring" microstates (not
necessarily in the Monte Carlo sample) with energy $E\pm\Delta E$ and
magnetization $M\pm\Delta M$, which can be reached from the particular
microstates of the sample by applying single spin flip operations. 
Again, from this extended histogram, the density of states can be computed
(cf.~Sec.~\ref{newsimobs}).

As already pointed out by Oliveira \cite{Oliveira3}, the advantage of the 
transition observable method is that every microstate of the 
Monte Carlo sample is investigated much more extensively than by
the standard  histogram technique.
Therefore, given a certain sample of microstates,
the density of states can be calculated more accurately from simulation
averages of the transition observable introduced below than by standard
methods. Additionally, as it is the selection of
microstates using pseudo random
numbers which is costly in computing time, the increase in computing time
from such a more extensive exploration of the chosen microstates is
absolutely negligible%
\footnote{%
At least in the case of the particular realization of the method introduced below.%
}%
. However, the effect on the data quality is significant and can amount to
orders of magnitude.
In the case of the examples studied in this paper, 
we find an efficiency gain
of roughly two orders of magnitude! This efficiency gain can
be expected to grow proportional to $L^{d/2}$, the square root of the volume of the system. 

It is this enormous gain of efficiency which should motivate
the reader to occupy himself with the underlying formalism, which is indeed simple
to implement in a simulation, but is somewhat heavy to formalize.\\ 
Section \ref{preliminaries} and \ref{general} aim 
to familiarize
the reader with the language used throughout this paper and with some
aspects of the Monte Carlo procedure. In section \ref{histotech}, the standard
histogram technique is reviewed in the context of the calculation of the 
density of states. The transition observable is introduced
in Sec.~\ref{newsimobs}. In Sec.~\ref{oliviera}, it is shown that this
observable includes the one presented in reference \cite{Oliveira2} as a 
special case.
The rest of the paper (section \ref{comparison}) is devoted to the comparison
of the efficiency of computer simulations using the new method in
contrast to a standard histogram technique. This is done for the examples of
$2d$- and $3d$-Ising systems where we find a speed-up factor of
$\approx 40$ in the $32^2$ Ising system and $\approx 250$ in the $10^3$ Ising
system.

\section{Calculating the density of states by Monte-Carlo
  simulation}
\label{omega_calcul}
\subsection{Conventions and notation}
\label{preliminaries}
In this paper, we use the language of discrete Ising systems on hypercubic lattices
of linear size $L$ in $d$ spatial dimensions
with Hamiltonian
\begin{eqnarray}
  \label{hamiltonian}
  {\cal H}(S) & := & -J\sum \limits_{\langle i,j \rangle} \sigma_i \sigma_j
  - h \sum \limits_{i}\sigma_i \nonumber \\
  & =: & E(S)-h M(S) \;, \qquad S\in \Gamma_{L^d} \; ,
\end{eqnarray}
where $h$ denotes an external magnetic field.
$E(S)$ is the interaction energy and $M(S)$ the mag\-ne\-tization of the particular microstate
$S=\sigma_1,\sigma_2,...,\sigma_i,...,\sigma_{L^d}$ 
($=$ particular configuration of the
spins $\sigma_i$, $i=1,2,..,L^d$ on the $L^d$ lattice) with $\sigma_i \in \{-1,+1\}$.
The configuration space of the Ising system is denoted by $\Gamma_{L^d}$, and $\langle i,j \rangle$ indicates a summation over all pairs of nearest neighbours.\\ 
The discreteness of the Ising systems gives
rise to a {\it minimal} energy and magnetization spacing, denoted by
$\Delta E$ and $\Delta M$, respectively. Summations over interaction energy $E$
(magnetization $M$) cover all
energy (magnetization) values accessible.\\
In general, in order to simplify the notation, system
size dependencies are not stated explicitely. In what follows, all energies
are measured in units of the Ising coupling constant $J$, all Temperatures in
units of $J/k_B$ ($k_B$: Boltzmann's constant).

\subsection{Some remarks on Monte-Carlo simulations}
\label{general}
For a Monte
Carlo simulation, a Markov process is set up on configuration space
$\Gamma_{L^d}$ with a certain problem adapted stationary
distribution $\hat{w}(S)$, which is assumed%
\footnote{%
It is straightforward to extend the formalism introduced in the following
sections to the case of a more general stationary distribution. The
restricting assumption $\hat{w}(S)=w(E(S),M(S))$ is made only for the sake of
notational simplicity.%
} 
to depend only on the interaction energy $E$ and the magnetization $M$
of the microstate $S$, i.e.~$\hat{w}(S)=w(E(S),M(S))$.  From the Markov chain
$\{S\}_{\cal{N}}$ of length ${\cal{N}}$ (which, at least in the limit of
infinitely long samples, is distributed according to $\hat{w}$) the {\em
  simulation average} of an arbitrary function $f:\,\,\Gamma_{L^d}\rightarrow
\Bbb{R}$ on configuration space is obtained:
\begin{equation}                                                                                                                 
  \label{simav}                                                                                                                  
  \left\langle f(S) \right\rangle_{sim,w}(\{S\}_{\cal N})\,:=\,                                                                  
  \frac{1}{\cal N} \, \sum \limits_{S\in\{S\}_{\cal N}} f(S)                                                                     
  \,\stackrel{{\cal N} \to \infty}{\longrightarrow}\,                                                                            
  \sum \limits_{S\in \Gamma_{L^d}} f(S)\,\hat{w}(S) \;.                                                                             
\end{equation}                                                                                                                   
Of course, the simulation average depends on the stationary distribution and,
unless the length of the Markov chain reaches infinity, on the particular
sample $\{S\}_ {\cal{N}}$.

\subsection{Standard histogram technique}
\label{histotech}
The histogram $\H3$, which is proportional to the number of microstates of the sample with
interaction energy $E$ and magnetization $M$, is given by the simulation average of the observable
$\D{E(S)}{E}\D{M(S)}{M}$: 
\begin{eqnarray}
  \label{histo}
  \H3 & = &
            \left\langle\D{E(S)}{E}\D{M(S)}{M}\right\rangle_{sim,w}(\{S\}_{\cal N}) \nonumber \\
      & = &
            \frac{1}{\cal N}\,\sum \limits_{S\in
            \{S\}_{\cal N}}\D{E(S)}{E}\D{M(S)}{M} \nonumber \\
      & \stackrel{{\cal N} \to \infty}{\longrightarrow} &
        \sum \limits_{S\in\Gamma_{L^d}}\D{E(S)}{E}\D{M(S)}{M}\,w(E(S),M(S))
            \nonumber \\
      & = & \Omega(E,M)\,w(E,M)
            \;.
\end{eqnarray}
Since the underlying stationary distribution $w$ is known (at least beside an irrelevant factor), the density of states is obtained as
\begin{eqnarray}
  \label{micpart}
  \O & := & \sum \limits_{S\in \Gamma_{L^d}} \D{E(S)}{E} \D{M(S)}{M}
  \nonumber \\
   & = & \left. \lim \limits_{{\cal N}\to \infty} \H3 \right/ w(E,M) \;,
\end{eqnarray}
or --- more realistically for a computer simulation --- at least an estimator
for $\Omega$ is obtained by omitting the limiting procedure
$\lim_{{\cal{N}}\to \infty}$.

\subsection{Transition observable method}
\label{newsimobs}
In this section it is shown that the density of states can be obtained from
simulation averages of certain transition observables defined below, which have
the advantageous feature that they enable the estimation of the density of
states in a much more efficient way than the standard histogram method does. 

As a preliminary step, let us define the {\em microcanonical average} of any
system observable $f(S)$:
\begin{eqnarray}
  \label{micav}
  \average{f(S)}(E,M) & := &
  \lim \limits_{{\cal N}\to\infty}\frac{\average{f(S)\,\D{E(S)}{E}\,\D{M(S)}{M}}_{sim,w}(\{S\}_{\cal N})}{\H3}
  \nonumber \\
  & = &
  \frac{\sum \limits_{S\in \Gamma_{L^d}}\D{E(S)}{E}\D{M(S)}{M}\,f(S)}{\O} \;
 .
\end{eqnarray}

Let ${\cal A}$ be a set of operators acting on configuration space
$\Gamma_{L^d}$
\begin{equation}
  \label{setAhere}
  {\cal A}\subseteq \left\{A:\,\, AS\in \konfig \,\,\, \forall \,\, S\in \konfig \right\} \; .
\end{equation}

The {\em transition observable} $\bhm{i}{j}(S)$ is defined as the number of operators $A\in{\cal{A}}$ acting on the particular microstate $S$, which result in microstates $\tilde{S}$ with interaction energy $E(\tilde{S})=E(S)+i\Delta E$ and magnetization $M(\tilde{S})=M(S)+j\Delta M$:
\begin{equation} \label{trans_obs2}
  \bhm{i}{j}(S)  := 
  \sum \limits_{\tilde{S}\in \Gamma_{L^d}} \D{E(\tilde{S})}{E(S)+i\cdot \Delta
  E}\,\D{M(\tilde{S})}{M(S)+j\cdot \Delta M} \sum \limits_{A\in {\cal A}}
  \D{AS}{\tilde{S}} \;; \qquad i,j\in {\mathbb Z} \; .
 \end{equation}
(See Fig.~\ref{visualization} for an illustration of the thus defined observables.)
Then, for any set of operators ${\cal A}$ which satisfies
\begin{equation}
  \label{bhmeqns}
  0 \,\ne\,
  \sum \limits_{S\in \konfig}\D{E(S)}{E}\,\D{M(S)}{M}\,\bhm{i}{j}(S) \,=\,
  \sum \limits_{S\in \konfig}\D{E(S)}{E+i\cdot \Delta E}\,\D{M(S)}{M+j\cdot
  \Delta M}\,\bhm{-i}{-j}(S) \; ,
\end{equation} 
the density of states $\Omega (E,M)$ can be calculated from the microcanonical
averages (\ref{micav}) of the thus defined transition observables to yield
\begin{equation}
  \label{bhmeqn}
  \Omega {\scriptstyle (E,M)} \,\,=\,\,
  {\textstyle
  \frac{\vphantom{\Big( \Big)}\micavbf{\bhm{-i}{-j}(S)}{E+i\Delta
  E}{M+j\Delta M}}{\vphantom{\Big( \Big)}\micavbf{\bhm{i}{j}(S)}{E}{M}}
  } 
  \,\,\Omega 
  {\scriptstyle (E+i\Delta E,M+j\Delta M)} \; .
\end{equation}
Remarks:
\begin{enumerate}
\item Microreversibility as explained in App.~\ref{microrev} is a sufficient condition
for the equality in (\ref{bhmeqns}). Apart from this condition which is implemented easily, 
the set ${\cal A}$ of operators can be chosen arbitrarily.
\item In the density of states, a multiplicative constant is physically
  irrelevant! For that reason, $\Omega $ can be chosen arbitrarily for one
  particular value of $(E,M)$.
  Then, the density of states of the remaining $(E,M)$
  values can be calculated from Eqn.~(\ref{bhmeqn}).
\item The efficiency of the transition observable method depends crucially on the
particular choice of ${\cal A}$.
\item The transition observable method is {\em neither} restricted to the
investigation of Ising systems (with bare next neighbour interaction, cf.~\cite{Lima}) {\em nor}
to the investigation of discrete
systems (cf.~\cite{Munoz}). Example: consider a discrete spin system with a
Hamiltonian consisting of two interaction terms
\begin{equation}
  {\cal H}(S)\,=\,E_1(S)+E_2(S) \; ,
\end{equation}
which depends on certain coupling constants, say, $J_1$ and $J_2$ (e.g.~ferromagnetic
coupling to next neighbours and antiferromagnetic coupling to next-nearest
neighbours).  
The knowledge of the density of states as a function of $E_1$ and $E_2$, i.e.
\begin{equation}
  \Omega(E_1,E_2)\,:=\,\sum \limits_{S\in\konfig}
  \D{E_1(S)}{E_1}\,\D{E_2(S)}{E_2} \; .
\end{equation}
enables the determination of the
thermostatic properties of the system for {\em all} possible values of the
ratio of the coupling constants by applying certain "skew-summing" techniques
(cf.~\cite{Deserno}). In complete analogy to the above,
a set of transition observables can be defined, which 
facilitates the determination 
of the thus defined density of states $\Omega(E_1,E_2)$. 
\item For a matrix $\mathbf{T}$ defined as
  \begin{equation}
    \label{TMMC}
    \left[ \mathbf{T} \right]_{(E^\prime,M^\prime),(E,M)}
     :=
    \frac{1}{|{\cal A}|}  \micavbf{\bhm{\frac{E^\prime-E}{\Delta
     E}}{\frac{M^\prime-M}{\Delta M}}(S)}{E}{M} \; ,
  \end{equation}
where $|{\cal A}|$ is the cardinality of the set ${\cal A}$, it is easy
to show that 
\begin{itemize}
\item[a)] $\mathbf{T}$ is a stochastic matrix, i.e.
\begin{equation}
 \left[ \mathbf{T}\right]_{(E^\prime,M^\prime),(E,M)} \ge 0      
 \quad  \forall \,\,(E^\prime,M^\prime),(E,M)
\end{equation}
and
\begin{equation} \label{stochastic}
\sum \limits_{(E^\prime,M^\prime)} \left[ \mathbf{T}
  \right]_{(E^\prime,M^\prime),(E,M)}=1
\end{equation}
\item[b)] The density of states is the stationary state of
  $\mathbf{T}$:
\begin{equation}\label{eigenstate}
\sum \limits_{(E,M)} \left[ \mathbf{T}
  \right]_{(E^\prime,M^\prime),(E,M)} \Omega (E,M) = \Omega
  (E^\prime,M^\prime) \; .
\end{equation}
\end{itemize}
Furthermore, if $\mathbf{T}$ is regular, i.e. if for all $E^\prime,M^\prime$
and $E,M$
\begin{equation}\label{regular}
\exists n \in {\mathbb{N}} : \left[ {\mathbf{T}}^m
\right]_{(E^\prime,M^\prime),(E,M)} > 0 \;\; \forall m \ge n \; ,
\end{equation}
the stationary state of $\mathbf{T}$ is unique. Nevertheless, even if this
condition is not fulfilled, the density of states can be computed piecewise on
certain subsets ${\cal E, M}$ of the total set of possible energy and
magnetization values of the system. The thus produced "fragments" of the density
of states are then connected to each other via (a priori unknown)
multiplicative constants.\\
Note here that the stochastic matrix $\mathbf{T}$ defined above is closely
related to the so--called Transition Matrix Monte Carlo method introduced by
Swendsen and Li \cite{SwendsenLi}.
\end{enumerate}

\subsection{Reduction to Oliveira's observable: the reduced
  transition observable method}
\label{oliviera}
The results of Sec.~\ref{newsimobs} can be simplified to those presented by
Oliveira in reference \cite{Oliveira2}, where no information on the magnetization of
the system is regarded.
Formally, this can be achieved by a summation over the magnetization $M$
(or the index $j$, respectively) in some of the expressions of the preceding section.
Then, however, only a determination of the {\em reduced density of states}
\begin{equation} \label{redomega}
  \Ored := \sum \limits_M \O
\end{equation}
is feasible, which does not entail the entire thermodynamic information of the
system (in the sense that $\Omega (E,M)$ enables the calculation of thermal and magnetic
equations of state in various ensembles whereas $\tilde{\Omega}(E)$ just
allows for the estimation of the thermal equation of state).\\
The {\em reduced microcanonical average} of any system
observable $f(S)$ over the energy--shell $E(S)=E$ is defined:
\begin{equation}
  \label{enav}
  \average{f(S)}(E) \,:=\,
  \lim \limits_{{\cal N}\to\infty}\frac{\average{f(S)\,\D{E(S)}{E}\,}_{sim,w}(\{S\}_{\cal N})}{\sum \limits_M\H3}
  \,=\,
  \frac{\sum
  \limits_{S\in\Gamma_{L^d}}\D{E(S)}{E}f(S)}{\Ored} \; .
\end{equation}
We further define the {\em reduced transition observable}
\begin{eqnarray}
\bhme{i}(S) & := & \sum_{j\in {\mathbb Z}} \bhm{i}{j}(S)\,=\nonumber\\
\label{trans_obs}
& = & \sum \limits_{\tilde{S}\in \Gamma_{L^d}}
\D{E(\tilde{S})}{E(S)+i\cdot \Delta E}
\sum \limits_{A\in {\cal A}}\D{AS}{\tilde{S}}
\;; \qquad i,j\in {\mathbb Z} \; .
\end{eqnarray}
Then, for any set of operators ${\cal A}$ which satisfies
\begin{equation}
  \label{iff}
  0 \,\ne \,
  \sum \limits_{S\in \Gamma_{L^d}} \D{E(S)}{E} \bhme{i}(S)
  \,=\,
  \sum \limits_{S\in \Gamma_{L^d}} \D{E(S)}{E+i\Delta E} \bhme{-i}(S)
  \; ,
\end{equation}
the reduced density of states (\ref{redomega}) can be calculated from the
reduced microcanonical average (\ref{enav}) of the reduced
transition observable (\ref{trans_obs}):
\begin{equation}
  \label{bhmeqn2}
\displaystyle
  \tilde{\Omega}{\scriptstyle (E+i\Delta E)}\,\,=\,\,
  \frac{\vphantom{\Big( \Big)}\scriptstyle\averagebf{\bhme{i}(S)}
  (E)}{\vphantom{\Big( \Big)}\scriptstyle\average{\bhme{-i}(S)}(E+i\Delta E)}
   \,\,\tilde{\Omega}{\scriptstyle (E)}\; .
\end{equation}
Again,
microreversibility (cf. appendix \ref{microrev}) is sufficient to ensure
the equality in (\ref{iff}).\\


\section{Comparison of the efficiency of the standard
histogram and the transition observable technique}
\label{comparison}
To demonstrate the advantages of the transition observable method, numerical
results obtained from either the standard histogram method or the transition
observable method are compared. The philosophy of the comparison is to use
some simulation technique to generate {\em one} sample of microstates
which then is evaluated according to {\em both} methods.\\
The simulations were performed for a $d=2$, $L=32$ and a $d=3$, $L=10$ Ising
system. 
For the sake of completeness, the details of the
computer simulations are given in App.~\ref{simulation}.\\
The set ${\cal A}$
of lattice operators was chosen to consist of $L^d$ operators, which are
labelled by the subscript $i$ and are defined by their action on a particular
microstate $S$:
\begin{equation}
  \label{action}
  A_i:\,\,\,
  S=\sigma_1,\sigma_2,...,\sigma_i,...,\sigma_{L^d}
  \,\,\mapsto \,\,
  \tilde{S}=\sigma_1,\sigma_2,...,-\sigma_i,...,\sigma_{L^d}
\end{equation}
i.e. the operator $A_i$ flips only the $i$-th spin of the Ising lattice.
Obviously, since $A_iA_iS=S$, the thus defined set of operators meets the
condition (\ref{requirement}) and therefore is microreversible.
Note that the determination of the simulation average of the transition observable by
use of this particular set of lattice operators can be done very
fast. In fact, the time needed for applying the $L^d$ operators of the set
${\cal A}$ to a particular
microstate is much shorter than the time needed to perform a lattice sweep!

Simulation averages of $\bhm{i}{j}$  were recorded only for values of
$i\in \{-1,0,1\}$ and $j\in \{-1,1\}$.

To emphasize the difference between the two methods, we compare "discrete
derivatives" (i.e.~ ratios of differences) of the logarithm of the density of
states, namely:
\begin{itemize}
\item[(i)] For the case of $d=2$ Ising model:
\begin{eqnarray} \label{betared}
  \Delta_E \left( \ln \tilde{\Omega}{\scriptstyle (E)}\right) & := &
  {\scriptstyle \frac{1}{2\Delta E}}\,\left[{\ln\tilde{\Omega}{\scriptstyle (E+\Delta
  E)}-\ln \tilde{\Omega}{\scriptstyle (E-\Delta E)}}
\right] \\
  & = &  {\scriptstyle \frac{1}{2\Delta E}}\,\ln 
  \left[
  {\textstyle
  \frac{\vphantom{\int \limits_a^a}  \averagebf{\bhme{1}(S)}(E-\Delta E)\,
                                     \averagebf{\bhme{1}(S)}(E)
  }{\vphantom{\int \limits_a^a}      \averagebf{\bhme{-1}(S)}(E+\Delta E)\,
                                     \averagebf{\bhme{-1}(S)}(E)}
  }
  \right] \; .\label{betabhm}
\end{eqnarray}
\item[(ii)] For the case of $d=3$ Ising model:
\begin{eqnarray}
  \label{htm} 
    \hspace{-6mm} 
  \Delta_M \left( \vphantom{\tilde{\Omega}}\ln \Omega {\scriptstyle (E,M)}\right)
  \hspace{-2mm}
  & := &
  {\textstyle \frac{1}{2\Delta M}}\,\,\left[\vphantom{\int \limits^a_a}
  {\textstyle \ln \Omega {\scriptstyle (E,M+\Delta M)}-\ln \Omega
  {\scriptstyle (E,M-\Delta M)}}\right]\\[1ex]
  & = & \hspace{-2mm}
   {\textstyle \frac{1}{2\Delta M}}\,
  \label{bhmeqn_ht}
  {\textstyle
  \ln \left[
       \frac{\vphantom{\int \limits_a^a}
             \micavbf{\bhm{1}{1}(S)}{E}{M-\Delta M}\,\,
             \micavbf{\bhm{-1}{1}(S)}{E+\Delta E}{M}
             }{\phantom{\int \limits_a^a}
             \micavbf{\bhm{1}{-1}(S)}{E}{M+\Delta M}\,\,
             \micavbf{\bhm{-1}{-1}(S)}{E+\Delta E}{M}}
       \right]}\\[1ex]
  \label{bhmeqn_ht2}
  \hspace{-2mm} & = & \hspace{-2mm}
{\textstyle \frac{1}{2\Delta M}}\,
  {\textstyle
  \ln \left[
       \frac{\vphantom{\int \limits_a^a}
            \micavbf{\bhm{-1}{1}(S)}{E}{M-\Delta M}\,\,
             \micavbf{\bhm{1}{1}(S)}{E-\Delta E}{M}
               }{\phantom{\int \limits_a^a}
             \micavbf{\bhm{-1}{-1}(S)}{E}{M+\Delta M}\,\,
             \micavbf{\bhm{1}{-1}(S)}{E-\Delta E}{M}}
       \right]}\\[1ex]
  \label{bhmeqn_ht3}
   \hspace{-2mm} & = & \hspace{-2mm}
   {\textstyle \frac{1}{2\Delta M}}\,   
  {\textstyle
  \ln \left[
       \frac{\vphantom{\int \limits_a^a}
             \micavbf{\bhm{0}{1}(S)}{E}{M-\Delta M}\,\,
             \micavbf{\bhm{0}{1}(S)}{E}{M}}{\phantom{\int \limits_a^a}
             \micavbf{\bhm{0}{-1}(S)}{E}{M+\Delta M}\,\,
             \micavbf{\bhm{0}{-1}(S)}{E}{M}}
       \right]} \; .
\end{eqnarray}
\end{itemize}
Note that both "discrete derivatives" (\ref{betared}) and (\ref{htm}) are
closely related to microcanonical equations of state (see
\cite{athens98,KPHpub,condmat} and appendix \ref{eqstate} for more details).

\subsection{Example 1: the $d=2$, $L=32$ Ising lattice}
\label{example1}
In Fig.~\ref{beta2d}, the differences of 
the logarithm of the reduced density of states
as emerging from the transition observable method (cf.~(\ref{betabhm}))
and the conventional
histogram method (cf.~(\ref{micpart}))
are shown together with the exact result \cite{Beale}. By use of a sample of $8\cdot 10^5$ microstates, the 
transition observable method yields a result which, on the scale of the
figure, can hardly be distinguished from the exact result, whereas the data
obtained from the histogram method scatter strongly around the latter.\\
In Fig.~\ref{diff2d}, the results of the transition observable
method (using {\it one} sample of $n=8\cdot 10^5$ microstates) are compared to the results of the
histogram method for several sample lengths ($n$, $5\cdot n$,
$10\cdot n$ and $15\cdot n$) by plotting the deviation of the
simulation data from the exact result. Even if the simulation
time is chosen $15$ times longer in the histogram method, the transition
observable method still yields more accurate results.\\
Calculating the mean square
deviation of the simulation data from the exact result as a function of
simulation time%
\footnote{%
The comparison of the data was done within a certain "energy window" which was
chosen around the centre of the histogram, i.e.~the tails of the histogram
have been discarded. Since the same sample is used in both evaluation
techniques, the result of the comparison does not depend on the width
of the "energy window" chosen for the evaluation of the $\chi^2$-deviations.
}
(Fig.~\ref{chi2_2d}), we notice that the accuracy of both methods is improved
according to a power law (the corresponding exponents seem to be the same
($\approx -1$) in both methods). But, at any given time, the transition observable method
beats the standard histogram method in accuracy by a factor of roughly
$40$. 

\subsection{Example 2: the $d=3$, $L=10$ Ising lattice}
\label{example2}
From the Monte Carlo samples, we computed 
the differences of the logarithm of the density of states in
direction of the magnetization
according to Eq.~(\ref{micpart}) in the case of the
histogram method and according to Eqs.~(\ref{bhmeqn_ht})-(\ref{bhmeqn_ht3}) 
in the case of the transition observable method (in fact, we computed the
mean value
of the three possibilities (\ref{bhmeqn_ht})-(\ref{bhmeqn_ht3}) of
determining $\Delta_M \left(\ln \Omega \right)$).

In Figs.~\ref{diff3d}a) and b), 
$\Delta_M \left(\ln \Omega \right)$ is shown for $E/10^3=-.924$. In
Fig.~\ref{diff3d}a), a sample
of length $10\cdot 10^6$ microstates is used for the evaluation
of $\Delta_M \left( \ln \Omega \right)$ according to both methods whereas in
Fig.~\ref{diff3d}b),
the histogram method with a sample length of $50\cdot 10^6$ microstates
is compared to the transition observable method with sample length
$10\cdot 10^6$ again. For a better visualization of the difference between
the two methods, an odd polynomial
($f_{fit}(M)=a\cdot M+b\cdot M^3+c\cdot M^5$)
was fitted to the transition observable data. Subtraction of the data of
Figs.~\ref{diff3d}a) and b) from this polynomial yields the plots shown in
Figs.~\ref{diff3d}c) and d). 
In the figures, the
transition observable (histogram) data are represented by the
solid lines (points). The plots of the differences show the consistency of
both methods, that is: both data sets scatter "randomly" around the fit
function. The data emerging from the transition observable method, however, are
much more accurate than the data emerging from the standard histogram method
even if much longer samples are used in the latter.

For a quantitative comparison between the two methods,
the mean square deviations of the simulation results, with respect to a fit%
\footnote{%
We have performed a weighted $\chi^2$ fit of an odd polynomial  
$ f_{fit}(M)=a\cdot M + b\cdot M^3+c\cdot M^5 $
to the data obtained
from a $50\cdot 10^6$ sample by applying the transition observable method.
The errors needed for the weighted fit have been produced by a jack-knife
blocking procedure using $25$ data sets of length $2\cdot 10^6$ sweeps.%
}
to
the best data obtained by the transition observable method, is shown as a
function of the simulation time in Fig.~\ref{chi2_3d}. The accuracy is 
improved according to a power law with exponent
$\approx -1$ in both methods. But: at any given time, the transition observable method
yields results which are more accurate than the results emerging from the
histogram method by a factor $\approx 1/250$ in the sense of the mean square deviation.
In order to produce results of similar quality,
the simulation time in the standard histogram method has to be $\approx 250$
times longer than in the transition observable method!

\subsection{General remarks on section \ref{comparison}}
\label{genremark}
\begin{enumerate}
\item The two examples discussed in the preceding sections show that a
simulation can be accelerated dramatically by use of the transition observable
method.
Here, it is not the algorithm
to speed up the simulation but it is the observable measured during the simulation!
The reason for this striking difference is indeed very
simple: while every microstate, which is decided to be part of the sample,
just yields one entry in a list in the histogram method, it might yield
many transitions to neighbouring states (neighbouring with respect to the
interaction energy and/or magnetization) and, hence, the statistics of the transition observable
method can be expected to be much better than the statistics of the
conventional histogram method. In fact, since the Ising systems under
consideration just allows for $5$ ($7$) different interaction energy changes
and only two magnetization changes%
\footnote{%
$\Delta E/J=(\pm 12),\,\pm 8,\,\pm 4,\,0$ and $\Delta M=\pm 2$ in $d=(3),\,2$.%
}
under single spin flip operations (in $d=2$ ($3$)), the set of operators
${\cal A}$ chosen above can be expected to shorten the computational effort by
a remarkable factor, roughly proportional to the square-root of the inverse
volume $L^{-d/2}$ of the system!

\item The change of interaction energy under a single spin
flip operation depends on the configuration of the spins in the very
neighbourhood of the particular spin to be flipped. The typical configurations
of neighbouring spins vary with the interaction energy of the whole system. For
that reason, the factor of proportionality of the efficiency gain 
in the sense of the $\chi^2$-comparison 
introduced above can be expected to depend on the mean interaction energy of the
histogram, which itself 
depends on the simulation parameters (i.e.~the stationary distribution).

\item The particular way of generating the sample of microstates
is not important in the context of the comparison of the two
methods introduced in Sec.~\ref{omega_calcul}. 
\end{enumerate}


\section{Conclusion}

A Monte Carlo simulation consists of two steps. The first step is the
generation of a sample or spot check of microstates. The second step is the investigation
of these microstates. Conventionally, if the aim is to speed-up the
simulation, the first step is modified while the second remains unchanged. We
have shown that a more extensive exploitation of the microstates of the
sample, i.e. taking simulation averages of the transition observable instead of just 
cumulating a standard histogram, can effectively speed-up the simulation by a
tremendous amount!

In an extremely straightforward implementation of the transition
observable me\-thod, we reach a speed-up which can be expected to be
proportional to the square root of the volume $L^{d}$ of the
system under consideration. Such a speed-up seems unattainable by an
improvement of the algorithm of the simulation, i.e. by modifying the first step
of the simulation.

Even though the transition observable method seems to be built for discrete
spin systems, one of us (J.D.~Munoz, cf.~\cite{Munoz}) has already shown
that the method can be transferred to continuous spin systems. 

An extension of this method to "non-spin" systems like polymers might be a topic of 
future investigations.


\acknowledgments
We would like to thank A.~H\"uller for valuable discussions
and P.M.C. de Oliveira for some remarks on the history of the
Broad Histogram Method cited in Refs.~\cite{Oliveira}, \cite{Oliveira2}
and \cite{Oliveira3}. 
One of us (J.D.~Mu\~noz) would like to 
thank H.J.~Herrmann for hospitality and the Deutscher Akademischer Austauschdienst 
for financial support through scholarship A/96/0390.

\appendix
  \section{Microreversibility}
\label{microrev}
Let ${\cal A}$ be a set of operators acting on configuration space
$\Gamma_{L^d}$
\begin{equation}
  \label{setA}
  {\cal A}\subseteq \left\{A:\,\, AS\in \konfig \,\,\, \forall \,\, S\in \konfig \right\}
\end{equation}
such that for all $A\in {\cal A}$, there exists a unique inverse operator
$B=A^{-1}\in {\cal A}$, i.e.
\begin{equation}
  \label{requirement}
  \forall A \in {\cal A} \,\,\,\exists ! \,\,\, B\in {\cal A}:\,\,
  BAS=S \;.
\end{equation}
Then, ${\cal A}$ is said to show {\it microreversibility}.\\
From the microreversibility of $\cal{A}$, it follows immediately that the number of operators $A\in {\cal A}$ which transforms $S$ into $\tilde{S}$ equals the number of operators $A\in {\cal A}$ which transform $\tilde{S}$ back into $S$, i.e.
\begin{equation}
  \label{micro}
  \sum \limits_{A\in {\cal A}} \D{AS}{\tilde{S}}
  \,=\,
  \sum \limits_{A\in {\cal A}} \D{S}{A\tilde{S}}\;.
\end{equation}
Using the definition of the transition observable (\ref{trans_obs2}), this can be shown to be equivalent to
\begin{equation} \label{bhmeqnsapp}
  \sum \limits_{S\in \konfig}\D{E(S)}{E}\,\D{M(S)}{M}\,\bhm{i}{j}(S) \,=\,
  \sum \limits_{S\in \konfig}\D{E(S)}{E+i\cdot \Delta E}\,\D{M(S)}{M+j\cdot
  \Delta M}\,\bhm{-i}{-j}(S)\;.
\end{equation}
That is,
the number of operations which transform microstates with interaction energy
$E$ and magnetization $M$ into microstates with $E+i\Delta E$ and $M+j\Delta
M$ by use of operators $A\in{\cal{A}}$ is identical to the number of
operations which transform ``backwards'', i.e. from states with $E+i\Delta E$
and $M+j\Delta M$ to those with interaction energy $E$ and magnetization $M$.

\section{Microcanonical Equations of State}
\label{eqstate}
As mentioned in Secs.~\ref{comparison}, the
differences of the logarithm of the density of states 
$\Delta_M \left( \ln \Omega \right)$ are related to the microcanonical magnetic
equation of state.
Indeed, equation (\ref{htm}) is the 
microcanonical magnetic equation of state in a discrete notation (appropriate for the
description of finite Ising systems), which converges in the thermodynamic
limit $L \to \infty$ towards the magnetic equation of state of the infinite system
\begin{equation}
-\frac{h}{T}(\varepsilon,m)\,=\,\frac{\partial}{\partial m} \lim_{L\to\infty}L^{-d}\ln\Omega(E,M,L^{-1})\;,
\end{equation}
where $\varepsilon:=L^{-d}E$ and $m:=L^{-d}M$ are intensive quantities. 
The difference of the logarithm of the reduced density of states, as defined
in Eq.~(\ref{betared}), converges towards $\beta(\varepsilon,h/T=0)$ of the
infinite system for zero external field and can serve to compute zero field
properties of the system%
\footnote{%
\mbox{$\beta(\e,h/T)$ is the derivative with respect to
$\e$ of the Legendre transform of} $\lim_{L\to\infty}L^{-d}\ln \Omega
(\e,m,L^{-1})$ with respect to $m$.%
}%
.\\
Note that it is unnecessary and a rather roundabout way to convert the thus
obtained data into the commonly used canonical quantities. For details on the
investigation of phase transitions in a microcanonical approach and a
microcanonical finite-size scaling theory see references \cite{athens98},
\cite{KPHpub} and \cite{condmat}.

\section{Details of the Monte Carlo simulation}
\label{simulation}
\subsection{Simulation of the $d=2$, $L=32$ Ising lattice}
\label{d2L32}
A Monte Carlo simulation of a $32^2$ Ising system with periodic
boundary  conditions was performed. The stationary
distribution of the underlying Markov-process was chosen to be proportional to
the Boltzmann weight $w(E(S),M(S))\propto {\rm exp}\left\{-{\cal H}(S)/T
\right\}$ with simulation parameters $h=0$ and $T=2.269$ (see
Sec.~\ref{preliminaries} for the
definition of the Ising Hamiltonian). We have
implemented a sequential lattice update with a "Metropolis-type" transition
rate $T(S\!\rightarrow \! S^\prime)=\min \{1,\hat{w}(S^\prime)/\hat{w}(S)\}$ 
and we have sampled every $L^2$
configuration only. 
After "equilibration" ($6.4\cdot 10^5$ lattice sweeps have been discarded), 
several successive samples of $8\cdot 10^5$ microstates were taken. 

\subsection{Simulation of the $d=3$, $L=10$ Ising lattice}
\label{d3L10}
A Monte Carlo simulation of a $10^3$-Ising
system with periodic boundary conditions was performed. 
The stationary distribution was chosen to be
$w(E(S),M(S))\propto \{(E_0-E(S))/N_0\}^{(N_0-2)/2}$, i.e. independent
of $M(S)$ again. The parameters were chosen to be 
$E_0=1586$
and 
$N_0=1000$ (for a detailed discussion and interpretation of this stationary
distribution, see \cite{HullGerling}). 
The way of updating the lattice configurations  
is the same as for the simulation of the
$32^2$-Ising system (cf.~App.~\ref{d2L32}). After "equilibration" ($2\cdot
10^6$ lattice sweeps have been discarded), several successive samples of
$2\cdot 10^6$ microstates were taken.


\phantom{oben}
\vspace{4cm}
\begin{center}
\stepcounter{figure}
\vspace{4cm}
\begin{figure}[h]
\setlength{\unitlength}{0.00087500in}%
\begingroup\makeatletter\ifx\SetFigFont\undefined%
\gdef\SetFigFont#1#2#3#4#5{%
  \reset@font\fontsize{#1}{#2pt}%
  \fontfamily{#3}\fontseries{#4}\fontshape{#5}%
  \selectfont}%
\fi\endgroup%
\begin{picture}(5649,2274)(2689,0)
{\put(5000,0){\bf Figure 1:}
\thicklines
\put(2701,-1186){\line( 0,-1){2250}}
\put(2701,-3436){\line( 1, 0){2250}}
\put(6076,-1186){\line( 0,-1){2250}}
\put(6076,-3436){\line( 1, 0){2250}}
\put(2925,-1361){\bf a)}
\put(6300,-1361){\bf b)}
\put(4276,-3515){\line( 0, 1){150}}
\put(4176,-3715){$E+\Delta E$}
\put(3376,-3515){\line( 0, 1){150}}
\put(2776,-3715){$E- \Delta E$}
\put(3826,-3515){\line( 0, 1){150}}
\put(3726,-3715){$E$}
\put(2625,-1861){\line( 1, 0){150}}
\put(1725,-1911){$M+ \Delta M$}
\put(2625,-2311){\line( 1, 0){150}}
\put(2325,-2361){$M$}
\put(2625,-2761){\line( 1, 0){150}}
\put(1725,-2811){$M-\Delta M$}
\put(7201,-3515){\line( 0, 1){150}}
\put(7101,-3715){$E$}
\put(7651,-3515){\line( 0, 1){150}}
\put(7551,-3715){$E+\Delta E$}
\put(6751,-3515){\line( 0, 1){150}}
\put(6151,-3715){$E- \Delta E$}
\put(6000,-1861){\line( 1, 0){150}}
\put(5100,-1911){$M+ \Delta M$}
\put(6000,-2311){\line( 1, 0){150}}
\put(5700,-2361){$M$}
\put(6000,-2761){\line( 1, 0){150}}
\put(5100,-2811){$M-\Delta M$}
}
%
{\linethickness{.05mm}
\put(4276,-2986){\line( 0, 1){1350}}
\put(3376,-2986){\line( 0, 1){1350}}
\put(3826,-2986){\line( 0, 1){1350}}
\put(3151,-2311){\line( 1, 0){1350}}
\put(3151,-2761){\line( 1, 0){1350}}
\put(3151,-1861){\line( 1, 0){1350}}
\put(6526,-1861){\line( 1, 0){1350}}
\put(6526,-2311){\line( 1, 0){1350}}
\put(6526,-2761){\line( 1, 0){1350}}
\put(7201,-2986){\line( 0, 1){1350}}
\put(7651,-2986){\line( 0, 1){1350}}
\put(6751,-2986){\line( 0, 1){1350}}
}
{\thicklines
\put(3826,-2311){\vector( 1, 1){450}}
\put(3826,-2311){\vector( 0, 1){450}}
\put(3826,-2311){\vector( 0,-1){450}}
\put(3826,-2311){\vector(-1, 1){450}}
\put(3826,-2311){\vector( 1,-1){450}}
\put(3826,-2311){\vector(-1,-1){450}}
\put(7201,-1861){\vector( 0, 1){  0}}
\put(7201,-1861){\vector( 0,-1){450}}
\put(7201,-2311){\vector( 0, 1){  0}}
\put(7201,-2311){\vector( 0,-1){450}}
\put(7201,-1861){\vector(-1, 1){  0}}
\put(7201,-1861){\vector( 1,-1){450}}
\put(6751,-2311){\vector(-1,-1){  0}}
\put(6751,-2311){\vector( 1, 1){450}}
\put(7201,-2761){\vector( 1,-1){  0}}
\put(7201,-2761){\vector(-1, 1){450}}
\put(7651,-2311){\vector( 1, 1){  0}}
\put(7651,-2311){\vector(-1,-1){450}}
}
\end{picture}
\end{figure}
\end{center}


%
\newpage
\phantom{oben}
\vspace{4cm}
\begin{center}
\stepcounter{figure}
{\bf Figure \arabic{figure}:}
\vspace{4cm}
\begin{figure}[h]
\input{beta2d}
\end{figure}
\end{center}
\newpage
\phantom{oben}
\vspace{4cm}
\begin{center}
\stepcounter{figure}
{\bf Figure \arabic{figure}:}
\vspace{4cm}
\begin{figure}[h]
\input{diff2d}
\end{figure}
\end{center}
\newpage
\phantom{oben}
\vspace{4cm}
\begin{center}
\stepcounter{figure}
{\bf Figure \arabic{figure}:}
\vspace{4cm}
\begin{figure}[h]
\input{chi2_2d}
\end{figure}
\end{center}
\newpage
\phantom{oben}
\vspace{4cm}
\begin{center}
\stepcounter{figure}
{\bf Figure \arabic{figure}:}
\vspace{4cm}
\begin{figure}[h]
\input{multiexample3d1}
\end{figure}
\end{center}
\newpage
\phantom{oben}
\vspace{4cm}
\begin{center}
\stepcounter{figure}
{\bf Figure \arabic{figure}:}
\vspace{4cm}
\begin{figure}[h]
\input{chi2_3d}
\end{figure}
\end{center}

\newpage
\begin{center}
{\bf Captions:}
\end{center}
\vspace{5mm}
\setcounter{figure}{0}

\begin{center}

\begin{figure}[ht]
{\caption{\label{visualization}%
a) Visualization of the transition variables $\bhm{i}{j}$ for the 
case $i\in \{-1,0,1\}$ and $j\in \{-1,1\}$. 
Given a {\it particular} microstate $S$ with interaction energy $E(S)=E$
and magnetization $M(S)=M$, $\bhm{i}{j}(S)$ gives the number of 
possibilities to reach any state $\tilde{S}$ with energy 
$E(\tilde{S})=E(S)+i\cdot \Delta E$ and magnetization
$M(\tilde{S})=M(S)+j\cdot \Delta M$ by applying the set of 
lattice operators ${\cal A}$ to the microstate $S$. 
The microcanonical average of the transition
observable $\bhm{i}{j}$ is proportional to the {\it total} number
of possibilities for the event that, given {\it any} state $S$ with energy
$E(S)=E$ and magnetization $M(S)=M$, any other state $\tilde{S}$
with energy $E(\tilde{S})=E(S)+i\cdot \Delta E$ and magnetization
$M(\tilde{S})=M(S)+j\cdot \Delta M$ is reached under the action 
of ${\cal A}$. Fig.~\ref{visualization}b) shows the "transition
paths" corresponding to the various microcanonical expectation values
contributing to the differences of the logarithm of the density of states
$\Delta_M \left( \ln \Omega \right)$
at the point $(E,M)$, as introduced in Sec.~\ref{comparison}. 
}}
\end{figure}
\vfill

\begin{figure}[ht]
{\caption{\label{beta2d}%
Intensive energy as a function of $\left[\Delta_E \left( \ln \tilde{\Omega} \right)\right]^{-1}$ of 
a $32^2$ square Ising
system. The solid line is the exact result
, the dashed line
corresponds to the transition observable method and the points to the
conventional histogram method. 
The same sample of $8\cdot 10^5$ microstates
was used to perform both evaluations. The energy is plotted against 
$\left[\Delta_E \left(\ln \tilde{\Omega}\right)\right]^{-1}$ because of its correspondence to a thermal 
equation of state $E(T)$; cf.~App.~\ref{eqstate}.
The strong fluctuations in both the high energy and the low energy region of
the figure are due to poor statistics in the tails of the histograms.
In the central region, the difference between the dashed and the full line is
smaller than the line thickness!
}}
\end{figure}
\vfill

\begin{figure}[ht]
{\caption{\label{diff2d}%
In order to point out the differences between the transition observable
method and the conventional histogram method and in order to show 
that the used estimators are indeed unbiased, the data emerging from 
the Monte Carlo simulation are subtracted from the exact result.
The transition observable data (histogram data) are represented by solid lines (points).
}}
\end{figure}
\vfill

\begin{figure}[ht]
{\caption{\label{chi2_2d}%
In order to judge the quality of the two methods,
the $\chi^2$-deviation of the simulation results from
the {\it exact} result is shown as a function of the simulation time
(in units of $8\cdot 10^5$ lattice sweeps). 
}}
\end{figure}
\vfill

\begin{figure}[ht]
{\caption{\label{diff3d}%
a) and b): differences of the logarithm of the density of states
$\Delta_M \left( \ln \Omega  \right)$ for $E/10^3=-.924$. 
In Fig.~a), a sample
of length $10\cdot 10^6$ microstates is used for the evaluation
of $\Delta_M \left( \ln \Omega \right)$ according to both methods whereas in Fig.~b),
the histogram method with a sample length of $50\cdot 10^6$ microstates
is compared to the transition observable method with sample length
$10\cdot 10^6$ again. For a better demonstration of the difference of
the two methods, the same data are subtracted from a fit function 
in Fig.~c) and d) (see text for the details of the fit).
In all figures, the
transition observable data (histogram data) are represented by the
solid lines (points).
}}
\end{figure}
\vfill

\begin{figure}[ht]
{\caption{\label{chi2_3d}%
In order to compare the quality of the results emerging from 
the histogram method to those emerging from the transition observable method, 
the mean square deviations of the simulation results with respect to a fit to
the best data obtained by the transition observable method is shown as a 
function of the simulation time (in units of $2\cdot 10^6$ lattice sweeps).
}}
\end{figure}

\end{center}


\begin{thebibliography}{99}
\bibitem{Oliveira} In the original work \cite{Oliveira2}, Oliveira presented the transition observable method together with a new simulation technique. Unfortunately, the part concerning the simulational aspects contains some errors which were suspected by Berg and Hansmann \cite{Berg}, and explicitly pointed out as well as corrected by Wang \cite{Wang}. In this paper, the transition observable method is formulated independent of (any) simulation technique.
\end{thebibliography}
\end{document}